\documentclass[10pt]{article}
\usepackage{multicol,ftnright,citesort}
\begin{document}

\title{Depletion induced demixing in polydisperse mixtures
of hard spheres}

\author{{\bf Richard P. Sear}\\
~\\
Department of Physics, University of Surrey\\
Guildford, Surrey GU2 5XH, United Kingdom\\
email: r.sear@surrey.ac.uk}

\date{\today}

\maketitle

\begin{abstract}
Polydisperse mixtures are those in which components with
a whole range of sizes are present.
It is shown that the fluid phase of polydisperse hard spheres is
thermodynamically unstable unless the density of large
spheres decreases at least exponentially as their size increases.
The instability is with respect to
the large spheres crystallising out into multiple solid phases.
\end{abstract}

\vspace*{0.1in}

\begin{multicols}{2}

Mixtures of hard spheres in which spheres with a wide range
of diameters are present are a good first model of emulsions.
Emulsions are suspensions of droplets of oil or fat in water;
milk is perhaps the most familiar
example. The droplets of an emulsion interact
via a short ranged repulsion, which is well represented by a hard-sphere
interaction. They are typically present with a wide range
of diameters: from 0.1 to a few micrometers
\cite{packer72,mcdonald,bibette91}.
Mixtures in which a continuous range of sizes are present
are termed polydisperse \cite{salacuse82}.
They are much less well understood than systems
which contain only one or two components.
For example, the phase behaviour of single component hard spheres
\cite{hoover68} has
been understood for thirty years:
the fluid phase is stable up the point where the spheres occupy
a little less than half the volume of the suspension,
there is then a first order transition to a solid.
In contrast there are no phase diagrams known for polydisperse hard spheres.
Below, we examine polydisperse spheres with particular emphasis
on the largest spheres. We show that unless their density
decreases at least exponentially with increasing size,
they crystallise out of the mixture at all densities.
The mixture is then never stable as a single fluid phase.
The crystallisation is driven by a 
depletion attraction \cite{asakura,vrij76}
between the large spheres, due to the smaller spheres.
Depletion-induced separation of the largest spheres
has been observed in emulsions
\cite{bibette91} but there the floating of the droplets to the surface
due to gravity complicates the situation.
Our demonstration applies to spheres at equilibrium.

Specifying a polydisperse mixture requires specifying
the number density of spheres of every size.
This is done with a distribution function $x(\sigma)$ \cite{salacuse82}.
The number density
of spheres with diameter $\sigma$ is then $\rho x(\sigma){\rm d}\sigma$,
where $\rho$ is the total number density of spheres.
Although our final result will apply to a whole class of distribution
functions we choose a specific function for definiteness and
because it is widely used to describe emulsions \cite{packer72,mcdonald}
and powders \cite{granqvist76}. The distribution is called the log-normal
distribution, and it is defined by
\begin{eqnarray}
x(\sigma)&=&\frac{1+w^2}{{\overline \sigma}\sqrt{2\pi\ln(1+w^2)}}
\times\nonumber\\
&&\exp\left(-
\frac{\left[\ln(\sigma/{\overline \sigma})+(3/2)\ln(1+w^2)\right]^2}
{2\ln(1+w^2)}
\right),
\label{distf}
\end{eqnarray}
where ${\overline\sigma}$ is the mean diameter and $w$ is the
standard deviation in units of ${\overline\sigma}$. Note that
for this distribution there is no upper limit on $w$;
its lower limit is zero and corresponds to a one component system.
In the canonical ensemble a polydisperse mixture of hard
spheres is completely specified by $x(\sigma)$ and the total
number density of spheres $\rho$; for hard spheres the
temperature is not a relevant variable.

In order to make progress in understanding a polydisperse
mixture of hard spheres with a broad distribution of sizes
we distinguish between the spheres with diameters close
to or less than the average diameter ${\overline\sigma}$
and spheres with much greater diameters.
Due to the large differences in size and number density of
these two sets of spheres we will treat them differently.
Not only is the number density of the large spheres
much less than that of the spheres with diameters
near ${\overline\sigma}$ but the fraction of the
fluid's volume they occupy is much less.
This is so because $x(\sigma)$, Eq. (\ref{distf}), decays
much faster than $\sigma^{-3}$ at large $\sigma$.

Consider the very large spheres of the distribution, those with diameters
$\sigma\gg{\overline \sigma}$.
These spheres are immersed in a `sea' of spheres much smaller
than themselves,
for each large sphere there are many spheres with diameters
of the same order as ${\overline \sigma}$ or smaller. These
smaller spheres induce an effective attraction between the large
spheres of the polydisperse mixture: the well-known
depletion attraction first described by Asakura and
Oosawa \cite{asakura}. This effect has been extensively
studied both theoretically
\cite{vrij76,gast83,dijkstra98,poon95,lekkerkerker95}
and in experiments on colloids which accurately
model (polydisperse) hard spheres
\cite{poon95,lekkerkerker95,kaplan94,dinsmore95,dinsmore96,imhof95}.

The depletion attraction is entropic in origin (it cannot have any other
origin as in hard spheres there is no energy of interaction
and so there is nothing but entropy). When two
spheres approach each other then the volumes they exclude to
the other spheres overlap. Thus the volume this pair of spheres
denies to the other spheres decreases and so the
volume available to the other spheres increases, increasing
their entropy.
This is particularly pronounced
for a pair of large spheres surrounded by many small spheres, then
when the large spheres touch the entropy of very many small
spheres increases.

For a pair of spheres of diameter $\sigma$ immersed in an ideal gas
of spheres all of diameter $\sigma'\ll \sigma$ the range of the depletion
attraction  is $\sigma+\sigma'$. The attraction increases from
zero when the surfaces of the large spheres are $\sigma'$ apart
to a maximum when the large spheres touch.
The strength of the effective attraction can be measured by its value
at contact divided by the thermal energy $kT$, $u$.
This is the increase in the entropy of the small spheres
of size $\sigma'$ when a widely separated pair of spheres of size $\sigma$
is brought into contact with each other.
It is given by \cite{asakura,vrij76}
\begin{equation}
u=-\rho_sv_{ov},
\label{udepl}
\end{equation}
where $\rho_s$ is the density of the small spheres.
Each large sphere excludes the smaller spheres from a spherical
volume of diameter $\sigma+\sigma'$, shown in Fig. 1 by
the thick lines around the spheres.
When two large spheres are touching, the
two volumes which they exclude to the small spheres
overlap. The volume of overlap of these two volumes is $v_{ov}$.
In Fig. 1 we see that this volume
is equal to that of two caps, each an end of a sphere of diameter
$\sigma+\sigma'$ and of height $\sigma'/2$. We are considering
the limit of small $\sigma'/\sigma$ and so the caps are very flat.
Then the height of one of the caps a distance $x$ from
a line drawn between the centres of the two large spheres
is $(\sigma'/2)(1-x^2/r^2)$ where $r$ is the radius of a cap
at its base; $r^2=\sigma\sigma'/2$. 
The total volume of the two caps
\begin{equation}
v_{ov}=2\int_0^r2\pi x \left(\frac{\sigma'}{2}\right)
\left(1-\frac{x^2}{r^2}\right){\rm d}x
=\frac{\pi}{4}\sigma\sigma'^2.
\label{vov}
\end{equation}

Now, for polydisperse spheres distributed according to
Eq. (\ref{distf}) the number density of spheres with diameter
$\sigma'$ is $\rho x(\sigma'){\rm d}\sigma'$ and so for a pair of spheres
of diameter $\sigma\gg \sigma_c$ the depletion attraction
due to spheres with diameters $\le\sigma_c$ is
\begin{equation}
u(\sigma)=-\frac{\pi}{4}\rho \sigma
\int_0^{\sigma_c}x(\sigma')\sigma'^2{\rm d}\sigma',
\label{depl}
\end{equation}
which depends on the cutoff $\sigma_c$. However, if $\sigma$ is
sufficiently large that ${\overline \sigma}\ll \sigma_c\ll \sigma$ then
the integral depends only weakly on $\sigma_c$ because
$x(\sigma')\sigma'^2$ is small for values of $\sigma'\ge \sigma_c$.
Indeed we can extend the upper limit of integration to infinity
without introducing a significant error. The fact that we can do
so justifies our splitting of the distribution into two parts.
Then we have
\begin{equation}
u(\sigma)=-\frac{3}{2}\frac{\eta}{(1+w^2)^2}
\frac{\sigma}{\overline \sigma}~~~~~\sigma\gg{\overline\sigma},
\label{depl2}
\end{equation}
where we have used the relation
$\eta=(\pi/6)\rho {\overline \sigma}^3(1+w^2)^3$ which holds when
$x(\sigma)$ is given by Eq. (\ref{distf}).
The physical content
of this approximation is that very large spheres only
notice spheres with diameters around ${\overline \sigma}$ and less;
the density of the larger spheres is too small to add
significantly to the depletion effect.
Note that the attraction
increases linearly with the size of the spheres $\sigma$.

By using the idea of a depletion attraction we have reduced our
polydisperse mixture to the large $\sigma$ tail of the distribution
interacting via an effective interaction which is the
sum of a hard-core interaction plus the short ranged attraction of
Eq. (\ref{depl2}).
The attraction of Eq. (\ref{depl2}) favours condensed phases where
the large spheres are within the range of the attraction of each other.
Competing against this attraction is the translational entropy
of the large spheres, which favours dilute phases. This competition is
the same as that involved in the vapour-liquid transition of
a simple substance such as argon.
The translational entropy of the
large spheres is just that of an ideal gas mixture,
so per large sphere of
size $\sigma$ it is \cite{salacuse82}
\begin{eqnarray}
s_F(\sigma)&=&1-\ln[\rho x(\sigma)]\\
&\sim& \mbox{const.}-
\ln(\rho/{\overline\sigma})\nonumber\\
&&
-(3/2)\ln(\sigma/{\overline \sigma})
+\frac{\left[\ln(\sigma/{\overline \sigma})\right]^2}
{2\ln(1+w^2)},
\label{sf}
\end{eqnarray}
where the second expression is obtained by substituting Eq.
(\ref{distf}) in the first and the constant is a function only of $w$.
The entropy $s_F$ increases with sphere size $\sigma$
because the density decreases. But it only increases
as the square of a log, which is a slower than linear increase.

The stability of the dilute fluid phase
with respect to condensation into a phase in which the density
of large spheres is much higher, is determined by the relative
entropy of the dilute and condensed phases.
Therefore, we require the entropy of the condensed phase.
The range of the depletion attraction is $\sim{\overline\sigma}$.
For spheres of diameter $\sigma\gg{\overline \sigma}$,
this is very small in comparison to the size of the sphere, $\sigma$.
In the $\sigma\rightarrow\infty$ limit, the ratio
of the range of the attraction to the size of the sphere,
${\overline\sigma/\sigma}$, tends to zero.
This is the sticky-sphere limit introduced by Baxter \cite{baxter68}.
Thus the large spheres are all near this sticky-sphere limit
and as $\sigma$ diverges the spheres tend towards the limit.
Stell has shown \cite{stell91,hemmer90} that
as the strength of the attractions is increased, a fluid of
sticky spheres does not condense to form a liquid but
collapses to form a close-packed solid, see
also Refs. \cite{bolhuis94,sear98}. So,
we look not for condensing of the large spheres into
a dense, liquid-like, phase but for collapse into a dense solid.
We therefore require the entropy in the dense
solid phase. By dense we mean sufficiently close to the close
packed density that the sphere is within the range of
the depletion attraction of its neighbours.

The entropy per large sphere, $s_K$ of a dense solid phase has two parts.
The first is the entropy associated with the motion of the large sphere,
and the second is the entropy gain of the small spheres when a large
sphere is brought close to twelve neighbouring spheres,
as it is in a dense face-centred-cubic or
hexagonal-close-packed lattice.
The first part is easily obtained from a cell theory \cite{sear98,buehler51}.
This assumes that each sphere is restricted to a cage formed
by its neighbours, which are taken to be fixed at their lattice positions.
For a solid with a lattice constant $a$, the centre of
mass of a sphere can move a distance $\sim(a-\sigma)$
from its lattice position without bumping into any of its neighbours.
The solid is formed due to attractive interactions so
the spheres must be close enough to each other to
attract each other throughout the cage which its neighbours form.
For this to be true $a$ must satisfy
$a < \sigma+{\overline\sigma}/2$,
and so $a-\sigma = c{\overline\sigma}$,
where $c$ is a parameter in the range $(0,1/2)$.
For monodisperse spheres the entropy is simply
the logarithm of the volume
$(c{\overline\sigma})^3$, which is available
to the centre of mass of a sphere \cite{buehler51}.

When the spheres are polydisperse the situation is more complicated.
There is an upper limit to the range of sizes of spheres a single
solid phase can tolerate \cite{sear98b,bartlett,pusey87,barrat86}.
A lattice can only accommodate spheres up to about its lattice constant
$a$ in diameter;
larger spheres cannot fit into the lattice position without
overlapping with their neighbours.
Spheres with diameters less than
$\sim(a-{\overline\sigma})$ are so small that
they cannot be within the range of the depletion attraction
of all their neighbours \cite{sear98}.
This means that the large spheres cannot all crystallise into
a single solid phase. In order to crystallise they first
fractionate into many fractions, each containing spheres with
only a narrow range
of diameters. The fractions can then
crystallise individually to produce separate solid phases,
each containing spheres of a different size.
The combined fractionation-and-crystallisation of
polydisperse sticky spheres is discussed in Ref. \cite{sear98}.
The range of diameters is
roughly $a-\sigma=c{\overline\sigma}$.
This width of distribution contributes an amount
$\simeq\ln(c{\overline\sigma})$ to the entropy of the solid
\cite{salacuse82,sear98}.

The parameter $c$ will be determined by
a competition between the depletion attraction tending
to reduce it and the motion of the large sphere which tends to increase it.
However, our results are not sensitive to the exact value
of $c$ and so we merely take it to be much less than one.
Then the depletion attraction is almost equal to its
value at contact, Eq. (\ref{depl2}), and the
gain in entropy of the small spheres per large sphere which
solidifies is closely equal to minus six times Eq. (\ref{depl2}).
Then, the entropy per large sphere of a solid phase
of large spheres of size $\sigma$
\begin{equation}
s_K(\sigma)\simeq 4\ln(c{\overline\sigma})-6u(\sigma)~~~~~
\sigma\gg{\overline\sigma},~~c\ll 1.
\label{sk}
\end{equation}

The solid phases are much denser than
the fluid phase and so can be formed without increasing the volume
occupied by the system. Thus if the solid phases have a higher entropy
than the dilute fluid, then the dilute fluid cannot be the
equilibrium phase, as the entropy can be increased at fixed volume by
forming the solid phases. So, we now compare the entropies
of the fluid and solid phases.
The entropy change per large sphere $\Delta s$, when spheres of average
diameter $\sigma$ separate out from an ideal gas to form
a crystal phase with a polydispersity of order
$c{\overline\sigma}$ is Eq. (\ref{sk}) minus Eq. (\ref{sf})
\begin{eqnarray}
\Delta s(\sigma)&\simeq& \mbox{const.}+
\ln\left( c^4 \rho{\overline\sigma}^3\right)
+(3/2)\ln(\sigma/{\overline\sigma})
\nonumber\\
&&-\frac{\left[\ln(\sigma/{\overline \sigma})\right]^2}
{2\ln(1+w^2)}
+9\frac{\eta}{(1+w^2)^2}\frac{\sigma}{\overline \sigma}
~~~~~\sigma\gg{\overline\sigma}.
\label{deltas}
\end{eqnarray}
When the volume fraction $\eta$ is non-zero,
this is positive for sufficiently large $\sigma$.
In fact it is positive for any $x(\sigma)$ which decreases
more slowly than exponentially with $\sigma$.
Therefore, {\em any} fluid phase of hard spheres with a
distribution which decreases more slowly
than exponentially is {\em unstable}
with respect to the largest spheres crystallising
into solid phases with narrow polydispersities.
A single solid phase can contain only a
narrow slice, of width a fraction of ${\overline\sigma}$,
of the original distribution $x(\sigma)$ but
spheres with diameters ranging from infinity down to some
large but finite limit crystallise.
Thus, an infinite number of solid phases form; each phase with a different
range of sphere sizes.
For the sake of clarity, when we say that the fluid phase
is unstable we mean that the solid phases have higher entropies.
The fluid phase will however be metastable, i.e., stable with
respect to infinitesimal perturbations.

Very recently, Cuesta \cite{cuesta} has shown that within
the Boublik-Mansoori-Carnahan-Starling-Leland (BMCSL) \cite{salacuse82}
approximation polydisperse hard spheres with a log-normal
distribution with a sufficiently large standard deviation $w$
have a spinodal. Warren has also found a spinodal
within the BMCSL approximation \cite{warren}.
A spinodal is where the fluid phase becomes unstable
with respect to an infinitesimal density and composition
fluctuation. The difference between Cuesta's result and ours
is probably due to one or both of two factors.
The first factor is the nature of the transition we have found.
It is very strongly first order and so the transition occurs much before the
spinodal. The second  factor is the accuracy of the BMCSL approximation.
It may be poor when there are spheres of widely different sizes
present \cite{coussaert98}.

In comparing our result with experiment
it should be remembered that in an emulsion there will be some
upper size limit, beyond which there are
essentially no particles. Obviously, the
number of phases which separate out is then not infinite.
In addition, at sufficiently low volume fractions
the fluid phase of the emulsion will be stable.
The fluid phase can be destabilised by adding small spheres
to the distribution, so increasing the strength of the depletion
attraction. For emulsions, micelles can be added and indeed
this is done in Bibette's \cite{bibette91} procedure
for fractionating emulsions.

To summarise, polydisperse hard spheres with a non-zero
volume fraction and distributed according to a
distribution function which decays more slowly than exponentially are
thermodynamically unstable. Spheres above
some lower size limit crystallise due to the depletion
attraction induced between them by the presence of the smaller spheres
of the distribution. We have not determined this lower limit but it is much
larger than the average size ${\overline\sigma}$.
The solid phase can only tolerate a very limited polydispersity
\cite{sear98,sear98b,pusey87,barrat86} and the range of spheres
which crystallise is from this lower limit to infinity.
Thus, the number of solid phases which form is infinite.
This seems surprising at first but in the large $\sigma$ tail of the
distribution the depletion attraction is increasing more rapidly than
the translational entropy in the fluid phase and
so there is no upper limit to the sizes of spheres
which crystallise. The lower limit to the sizes of spheres
which crystallise is clearly finite as a solid phase
will form so long as it reduces the entropy by any non-zero amount.
The sub-linear increase with $\sigma$ of the translational entropy
in the fluid is the crucial factor in destabilising the fluid
phase. It inevitably leads to the fluid being unstable when
the attractions grow linearly with sphere diameter.

Finally, we conjecture that the instability we have found is
not restricted to spheres
or to attractions which arise from depletion.
Consider a general polydisperse fluid with a
number density $\rho x(l){\rm d}l$ of elements of size $l$
then the translational entropy increases as minus the logarithm of
$x(l)$. If the attractive energy (over $kT$)
$u(l)$ between elements increases faster than $\ln x(l)$
\begin{equation}
\frac{{\rm d} |u(l)|}{{\rm d} l}>-
\frac{{\rm d} (\ln x(l))}{{\rm d} l},
\label{gen}
\end{equation}
then the attractive energy
is much larger than the translational entropy for sufficiently large $l$.
When this is true we expect the mixture to be unstable with
respect to the largest elements condensing out to form a dense
phase in order to minimise the energy.

It is with pleasure that I acknowledge discussions with J. Cuesta.

%\newpage

\section*{Figure caption}

\noindent
Fig. 1. A schematic of two large touching spheres
of diameter $\sigma$, the shaded discs, with the volumes they exclude
to smaller spheres of diameter $\sigma'$. These
volumes are outlined by the heavy circles and they overlap
when the large spheres touch.

\end{multicols}

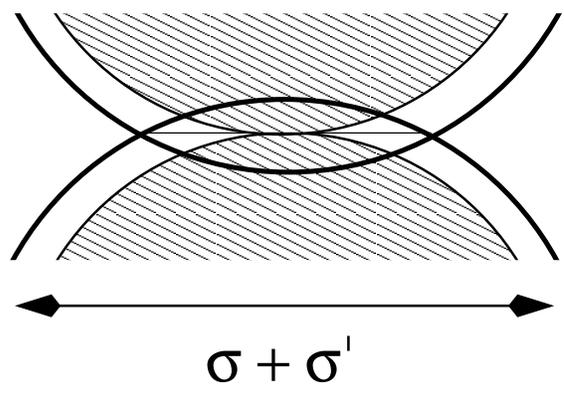
\end{document}